\documentclass[aps,prl,twocolumn,superscriptaddress]{revtex4}
\usepackage{amsmath}
\usepackage{amssymb}
\usepackage{graphicx}
\usepackage{wrapfig}
\usepackage{subfigure}
\usepackage{psfrag}

\begin{document}
\DeclareGraphicsExtensions{.eps}
\title{Topological spin liquid on the hyper-kagome lattice of Na$_4$Ir$_3$O$_8$}
\author{Michael J. Lawler}
\author{Hae-Young Kee}
\author{Yong Baek Kim}
\affiliation{Department of Physics, University of Toronto, 60 St. George St., Toronto, Ontario, M5S 1A7, Canada}
\author{Ashvin Vishwanath}
\affiliation{Department of Physics, University of California, Berkeley, CA, 94720}

\date{\today}

\begin{abstract}
Recent experiments on the ``hyper-kagome" lattice system Na$_4$Ir$_3$O$_8$ have demonstrated that it is a rare example of a three dimensional spin-$1/2$ frustrated antiferromagnet. We investigate the role of quantum fluctuations as the primary mechanism lifting the macroscopic degeneracy inherited by classical spins on this lattice. In the semi-classical limit we predict, based on large-$N$ calculations, that an unusual $\vec q=0$ coplaner magnetically ordered ground state is stabilized with no local ``weather vane" modes. This phase melts in the quantum limit and a gapped topological Z$_2$ spin liquid phase emerges. In the vicinity of this quantum phase transition, we study the dynamic spin structure factor and comment on the relevance of our results for future neutron scattering experiments.
\end{abstract}

\maketitle
\emph{Introduction.}
Geometrically frustrated spin systems are a promising place to search for new ``exotic" phases of matter. Classical spin models on such lattices typically exhibit massive degeneracy of their zero temperature ground state ensembles so that phases of matter in these systems are selected exclusively by thermal and/or quantum fluctuations. While magnetically ordered states may still arise at very low temperatures in such systems, emergent phases with, for example, fractionalized excitations and topologically protected ground states, are also possible, especially in spin $\frac{1}{2}$ systems\cite{Lhuillier:2002lr,Moessner:2001mx}.

\begin{figure}[t]
\psfrag{l}{$\ell$}
\psfrag{h}{$h$}
\subfigure[NIO's hyper-kagome lattice]{\includegraphics[width=0.23\textwidth]{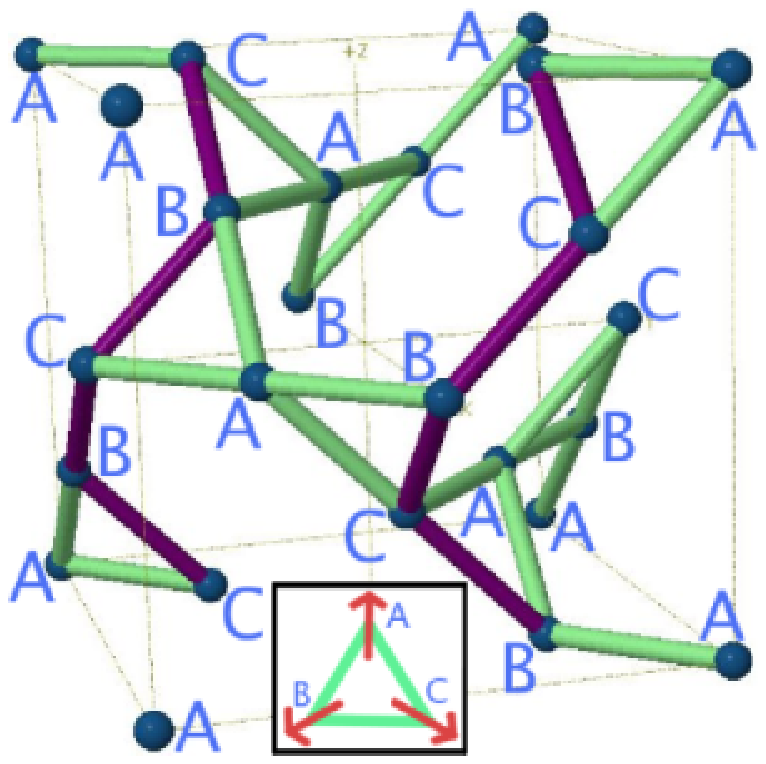}}
\subfigure[$h0\ell$-plane static structure factor]{
\begin{minipage}[b]{0.2\textwidth}
\includegraphics[width=\textwidth]{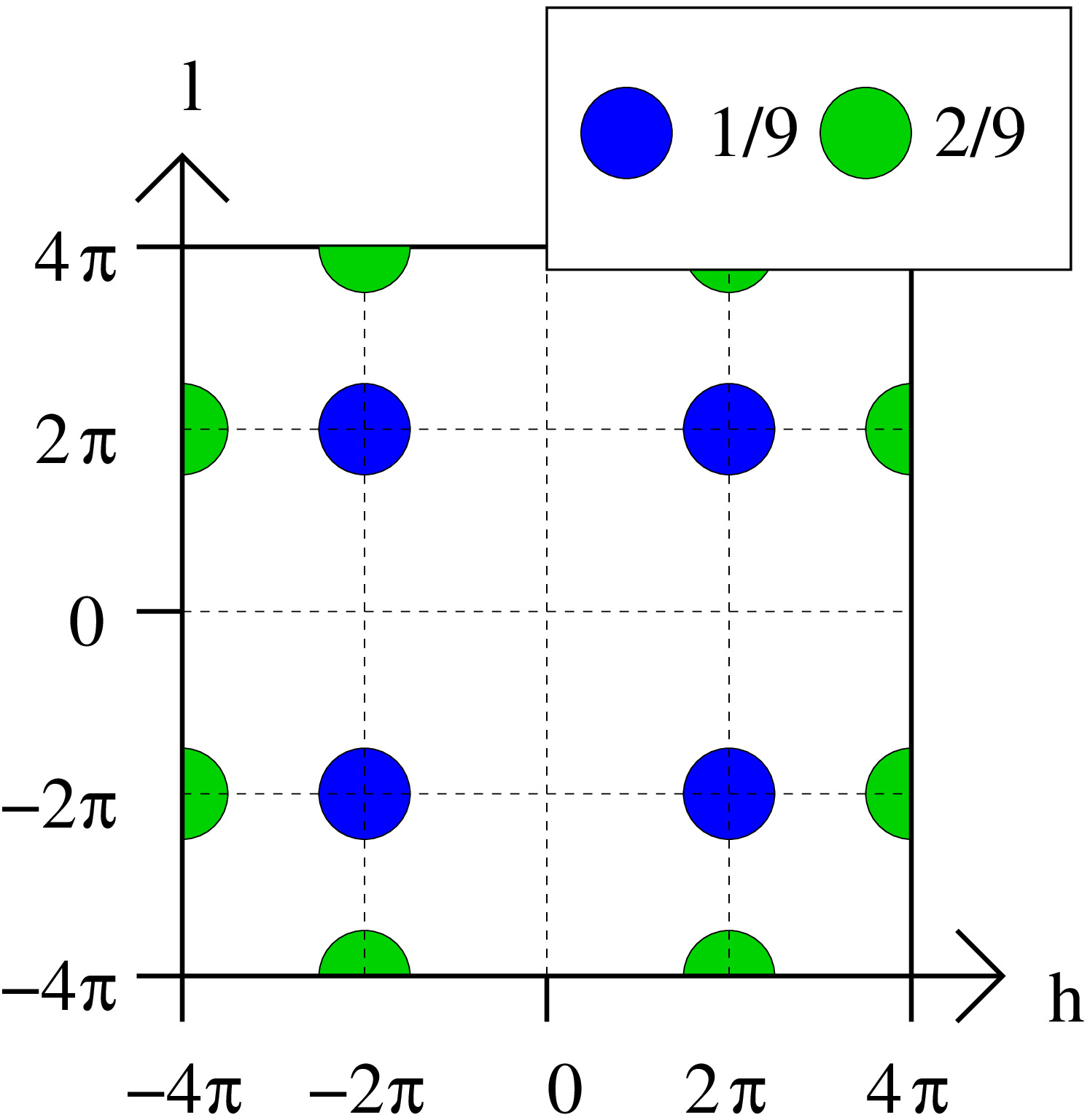}\\\ 
\end{minipage}}
\caption{(a) The unit cell of the iridium atoms in NIO. All sites and bonds are chemically equivalent due to the P4$_1$32 space group symmetry of the lattice. Letters indicate the 120$^o$ co-planer magnetic ordering expected in the semi-classical limit. (b) Positions of magnetic Bragg peaks in the $h0\ell$-plane of the ordered phase shown in (a). Notice magnetic Bragg peaks for this phase coincide with some lattice Bragg peaks.}
\label{fig:hk-unitcell}
\end{figure}

On the other hand, spin $1/2$ frustrated magnets are extremely rare, with only a few known examples such as Volborthite\cite{Hiroi:2001ej}, Herbertsmithite\cite{Helton:2007wx} and certain organic salts\cite{Shimizu:2003eo, Tamura:2006dp}. In light of this, the recent experiment\cite{Okamoto:2006fk} demonstrating that Na$_4$Ir$_3$O$_8$ (NIO) is a new spin $1/2$ geometrically frustrated magnet and the first such material to contain a genuinely three-dimensional lattice is especially important. In particular, these experiments see no sign of magnetic or orbital ordering down to a few Kelvin while bulk susceptibility measurements show that the spin $1/2$ iridium atoms interact anti-ferromagnetically with a Curie-Weiss constant of $650$ K. Furthermore a nearly constant susceptibility and a $C/T$ that is magnetic field independent with a broad anomalous peak at about $25$K is observed. Certainly, the remarkable hyper-kagome lattice (see Fig. \ref{fig:hk-unitcell}) that spin $1/2$ iridium atoms occupy is responsible for these striking observations.

In order to investigate whether NIO's magnetic properties are  truly quantum mechanical in origin, it is necessary to understand how the classical ground state degeneracy is lifted by both thermal fluctuations (entropic selection) and quantum fluctuations (energetic selection). Fortunately, much is now known about the entropic selection of these classical ground states due to a recent study of the classical Heisenberg model on NIO's hyper-kagome lattice\cite{Hopkinson:2006qm}. In particular, the classical cooperative paramagnetic (CCP) phase exists over a wide temperature range with a spin nematic phase setting in below $J/1000 \approx 300$ mK, where $J$ is the Heisenberg exchange coupling. However, the behavior of the classical model deviates from measurements of NIO's thermodynamic quantities at and below the anomalous peak in $C/T$ at 25 K suggesting that quantum fluctuations may dominate even at relatively high temperatures.

In this letter, we present litmus tests for the existence of a quantum spin liquid phase in NIO. We do so through an investigation of the ground states of the Heisenberg quantum antiferromagnetic spin model on its frustrated hyper-kagome lattice with the large-$N$ Sp($N$) method\cite{Read:1991fk, Sachdev:1991uq, Sachdev:1992qy}. This allows us to study the semi-classical (large ``spin") and quantum (small ``spin") regimes on equal footing. In particular, we show how a quantum spin liquid phase in the quantum regime can be distinguished from both a CCP phase at finite temperatures and the most likely magnetically ordered phase.

In the semi-classical regime, we predict a ground state with the 120$^o$ co-planer magnetic ordering shown in Fig. \ref{fig:hk-unitcell}a. Defining ABC as three unit spin vectors whose sum vanishes, this ordering is best characterized by a BCBC pattern along the ``threads", paths shown by the darker bonds. This is in contrast to the $\sqrt{3}\times\sqrt{3}$ state on the two-dimensional kagome lattice which is also co-planer but with a BCBC pattern around hexagons. Since threads are as long as the length of the system, this state is not expected to be entropically selected\cite{Chalker:1992fk}, a result consistent with the thermally stabilized spin nematic ordering and an absence of magnetic ordering found in the Monte Carlo study of Ref. \onlinecite{Hopkinson:2006qm}. Fig. \ref{fig:hk-unitcell}b shows the position of the corresponding magnetic Bragg peaks.

\begin{figure}[t]
\subfigure[$Z_2$ spin liquid hhl-plane (energy integrated)]{\includegraphics[width=0.22\textwidth]{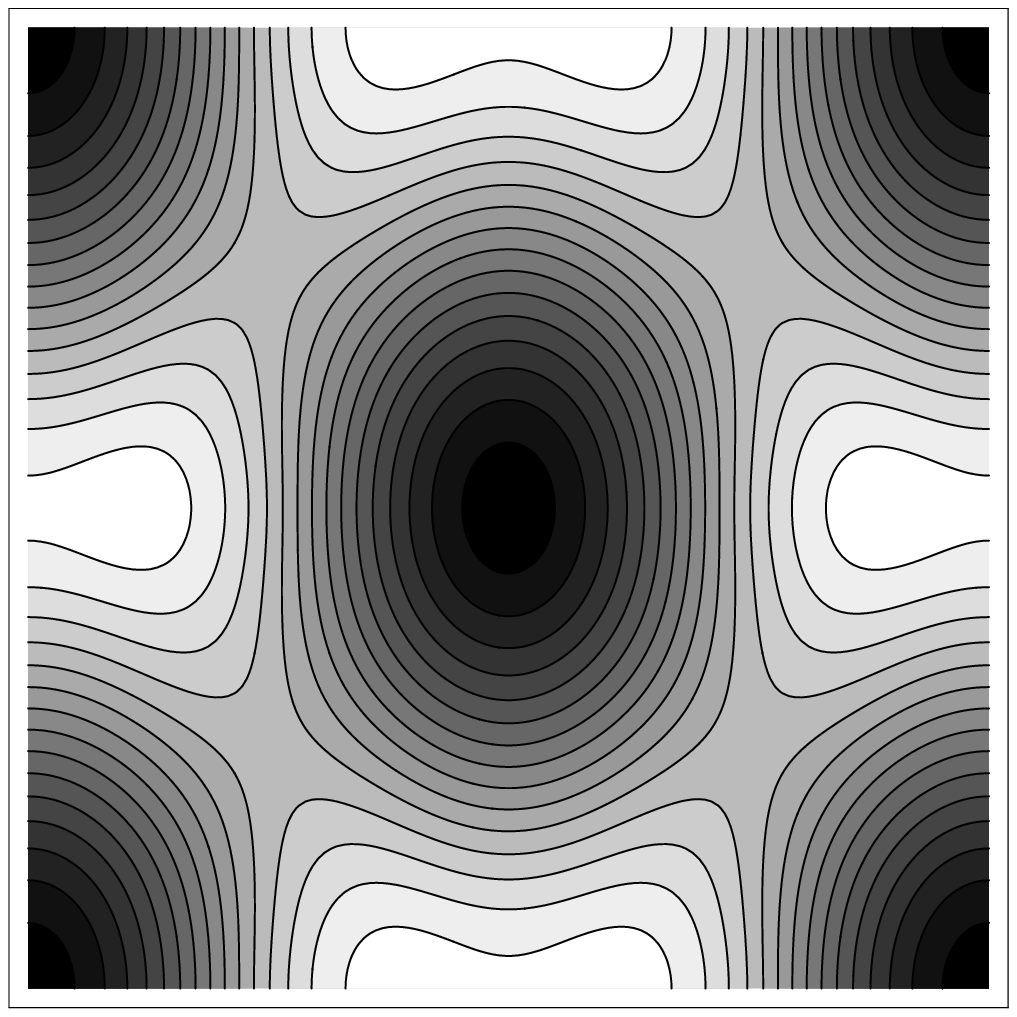}}
\subfigure[CCP phase hhl-plane\cite{Hopkinson:2006qm}]{\includegraphics[width=0.22\textwidth]{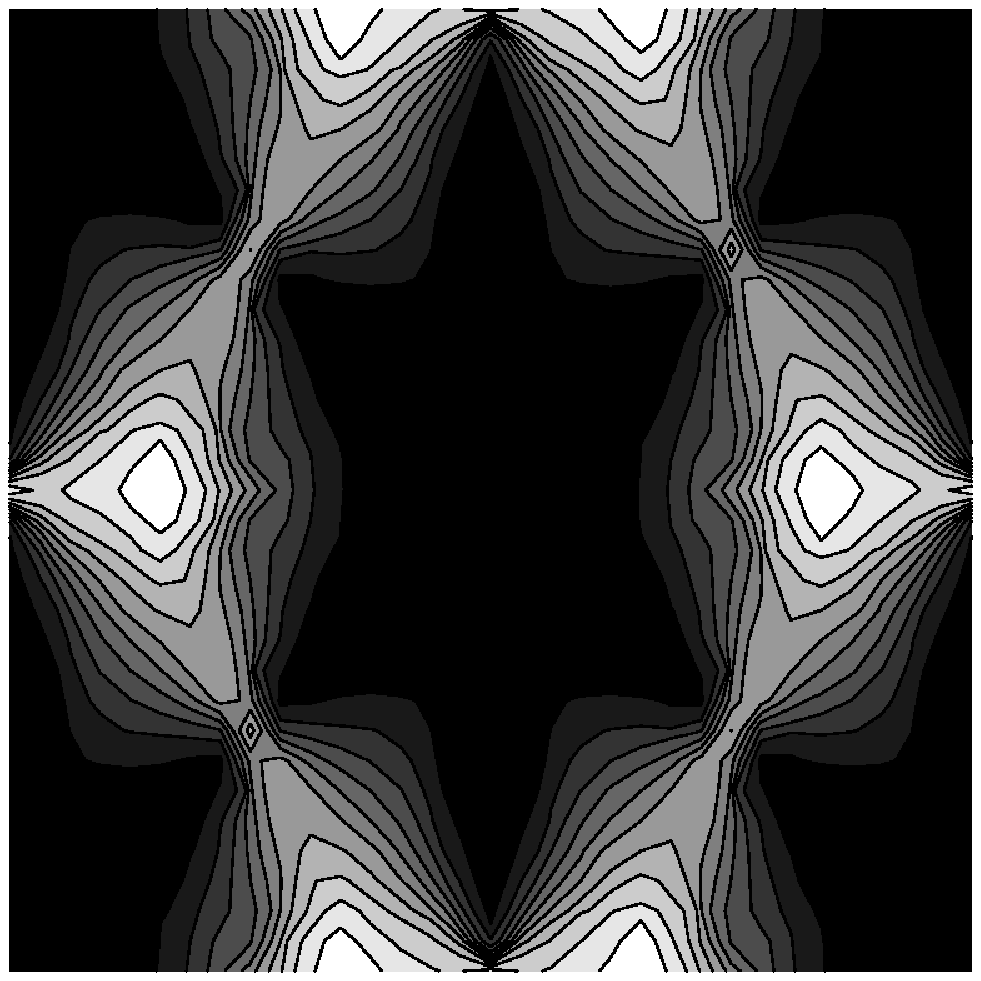}}
\subfigure[hhh$\omega$-plane (two-spinon threshold)]{\includegraphics[width=0.35\textwidth]{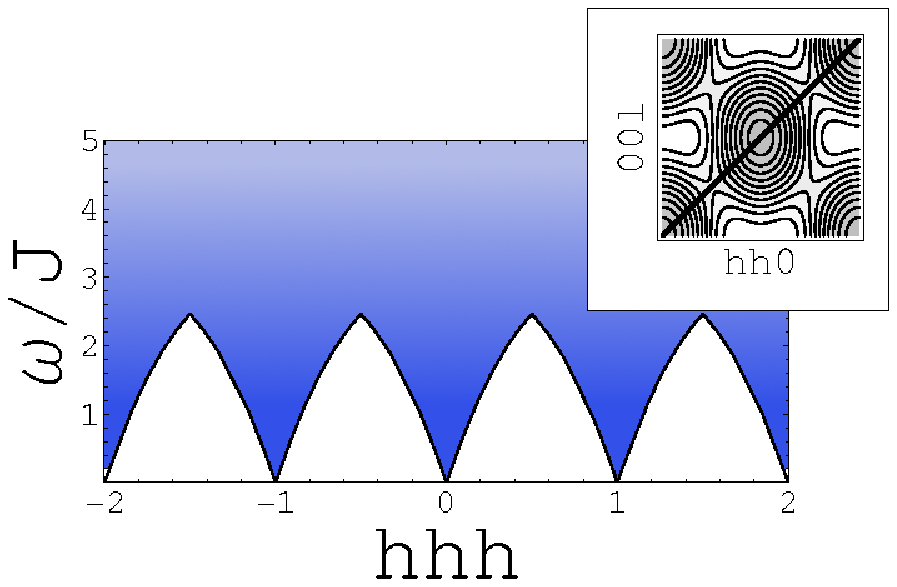}}
\caption{Distinguishing the $Z_2$ quantum spin liquid ground state from the classical cooperative paramagnet (CCP) via the dynamic spin structure factor ${S( k, \omega)}$. (a) and (b): the absence of the dipolar correlations in (b) lead to finite spin correlations along the diagonal [hhh] direction in (a) (axes run from $h=-2\ldots 2$, $l=-2\ldots2$). (c) the spinon-antispinon continuum in $S(k,\omega)$: shaded (blue) region represents where the structure factor is finite. Notice that the threshold of the spinon-antispinon continuum has a well defined dispersion.}
\label{fig:hkSk}
\end{figure}

On the other hand, in the quantum regime the magnetic ordering melts and a $Z_2$ quantum spin liquid phase arises. Since the spin liquid is a paramagnetic phase, it is important to distinguish it from the CCP phase. To this end, consider the dynamic spin structure factors $S(\vec k,\omega)$ depicted in Fig. \ref{fig:hkSk}.  The energy integrated spin structure factor of the $Z_2$ spin liquid ground state in Fig. \ref{fig:hkSk}a can be compared directly with the static spin structure factor of the CCP phase in Fig. \ref{fig:hkSk}b\cite{Hopkinson:2006qm}. In the CCP phase, long range dipolar correlations lead to the vanishing of spin correlations (nodes) along the $[hhh]$ and symmetry related directions. In contrast, dipolar correlations in the $Z_2$ spin liquid phase retain a finite correlation length, leading to the absence of this nodal structure.

Another important signature of the quantum spin liquid phase is the presence of spin-1/2 neutral excitations, dubbed spinons. This leads to a spinon-antispinon continuum in the dynamic spin structure factor and the bottom (or threshold in energy for a given momentum) of this continuum should have a well defined dispersion relation. Conversely, in a CCP phase no such sharp threshold is possible making its observation away from Bragg peaks a definitive proof for the existence of a quantum spin liquid phase. This threshold for the $Z_2$ quantum spin liquid phase proposed here is plotted in Fig. \ref{fig:hkSk}c (the shaded region simply represents where $S(\vec k,\omega)\neq 0$). 

\emph{The model.}
Consider now the Heisenberg model on NIO's hyper-kagome lattice. This model in the Schwinger boson representation is as follows\cite{Auerbach:1994wj}:
\begin{equation}
  H = J \sum_{\langle ij\rangle} \vec S_{i}\cdot \vec S_{j} = -\frac{J}{2}\sum_{\langle ij\rangle}\left[
   \mathcal{A}_{ij}^\dagger \mathcal{A}_{ij}^{ } - 2S^2\right]
\end{equation}
where the spin operators are $\vec S_{i}=\frac{1}{2}b_{i\sigma}\vec \tau_{\sigma,\sigma'}b_{i\sigma'}$, the singlet creation operator is $A^{\dagger}_{ij}=b^{\dagger}_{i\uparrow}b^{\dagger}_{j\downarrow}-b^{\dagger}_{i\downarrow}b^{\dagger}_{j\uparrow}$ and the constraint $\sum_{\sigma}b^\dagger_{i\sigma}b^{ }_{i\sigma}\equiv n_i=2S$ is imposed on each site, where $S$ is the size of the spin. It is convenient to generalize this SU(2) model by introducing $N$ flavors of Schwinger bosons, $b_{i\sigma}^m$, $m\in\{1,\ldots, N\}$ and letting $A^\dagger_{ij}\to\sum_{m=1}^{N}A_{ij}^{\dagger m}$ create a singlet for each flavor. Then %resulting Hamiltonian is then
\begin{multline}
H_{\text{Sp($N$)}} = J\sum_{\langle ij \rangle}\left[ \frac{N}{2}|Q_{ij}|^2\!-\!
  \sum_{m=1}^{N}\left(Q_{ij}^{ }\mathcal{A}^{\dagger m}_{ij} - Q_{ij}^*\mathcal{A}^{m}_{ij}\right) \right]\\
  + \sum_i \lambda_i\left(\sum_{m=1}^N n_{i}^m - N\kappa\right)
\end{multline}
and is Sp($N$) symmetric. In the large-$N$ limit $Q_{ij} = \sum_{m=1}^{N}\langle \mathcal{A}_{ij}^{m}\rangle$ is a mean field and the chemical potential $\lambda_i=\lambda$ enforces the constraint on average where $\kappa=2S/N$ is held fixed as $N\to\infty$.

This large-$N$ limit is well suited to our purposes since as a function of $\kappa$ we can access both a quantum regime at small $\kappa$ and a semi-classical regime at large $\kappa$. 

\emph{Quantum regime.} To study this regime, it is useful to start from an extreme quantum limit and  expand the ground state energy in powers of $\kappa$.  Following Ref. \cite{Tchernyshyov:2006lr}, before expanding in powers of $\kappa$, consider first reformulating the ground state energy in matrix form
\begin{equation}
\frac{E_{\text{Sp($N$)}}}{N}\! =\! \text{Tr}\bigg[\frac{J}{4} {\bf Q}\cdot{\bf Q}^\dagger\! -\! \lambda (1\!+\!\kappa){\bf I}\!
 +\! \lambda\sqrt{{\bf I}\!-\!\frac{J^2}{4\lambda^2} {\bf Q}\cdot{\bf Q}^\dagger}\bigg]
\end{equation}
where $({\bf Q})_{ij} = Q_{ij}$ and then re-scaling $Q_{ij}\to\alpha Q_{ij}$ and $\lambda\to\alpha\lambda$ by an $\alpha$ chosen so that $\partial E_{\text{Sp($N$)}}/\partial\alpha = 0$ making $E_{\text{Sp($N$)}}$ scale invariant. Solving the constraint equation, $\partial E_{\text{Sp($N$)}}/\partial\lambda = 0$, order by order in $\kappa$ then leads to
\begin{multline}\label{eq:Eloop}
  E_{\text{Sp($N$)}}/NN_{sites} = -\sum_{n=1}^{\infty}\bigg[
    2^{n-1}\frac{(2n-3)!!}{(2n)!!} \frac{P_n}{\left(P_1\right)^n} \\
    \pm \left(\text{terms involving} \frac{P_m}{\left(P_1\right)^m}, m < n\right)\bigg] \kappa^n
\end{multline}
where $P_n \equiv \text{Tr}[{\bf Q}\cdot{\bf Q}^\dagger]^n/N_{sites}$ is a measure of the total field strength, or flux (defined below), through all loops of length $2n$ starting and ending at the same site (including smaller loops whose total length is $2n$).  Through any individual loop $\mathcal{C}$ this (normalized) flux is defined by
\begin{equation}\label{eq:flux}
  2\cos\left( \Phi_{\mathcal {C}} \right) = Q^{ }_{ij} (-Q_{jk}^*) \ldots Q^{ }_{mn}(-Q^*_{ni}) +
  \text{c.c.}
\end{equation}
Thus, $P_n$ is the total flux through all loops of length $2n$.
%Incidentally, the coefficients of $P_{n}$ in Eq. \eqref{eq:Eloop} are the same as those obtained in a loop expansion of $E_{\text{Sp($N$)}}$ in the large $\kappa$ limit\cite{Hizi:2005qy}.

It follows from Eq. \eqref{eq:Eloop} that the ground state must satisfy a flux expulsion principle: in this state the flux $\Phi_{\mathcal{C}}$ through all loops must vanish, if possible. To see that this is the case, consider using Eq. \eqref{eq:Eloop} to study the energy difference between two states that differ only by flux through loops of length $\ell$ or greater
\begin{multline}\label{eq:deltaE}
  E_{\text{Sp($N$)}}(\Phi_\ell^{(1)}) - E_{\text{Sp($N$)}}(\Phi_\ell^{(2)}) = \\
      2^{\ell-1}\frac{(2\ell-3)!!}{(2\ell)!!} \frac{P_\ell^{(\Phi_\ell^{(2)})}-P_\ell^{(\Phi_\ell^{(1)})}}{P_1^\ell}\kappa^\ell + O(\kappa^{\ell+1})
\end{multline}
So, for example, if there is a  state with zero flux, $P_n$ is maximized and it is the ground state. For a nearest neighbor model, this result proves a previously made conjecture\cite{Tchernyshyov:2006lr}. Interestingly, as a by-product of Eq. \eqref{eq:deltaE}, we may also study the energetics of topological sectors that are characterized by loops that go all the way around the system (as discussed below).

\emph{Enumeration of spin liquid states.} In the quantum regime, the natural ground state may be a quantum spin liquid phase. Utilizing the above loop expansion, we can solve the energy minimization problem by first reducing the number of candidate ground states to a few spin liquid states and then comparing their energy. 

We begin by searching for spin liquid states among the set of translationally invariant Ising states with $Q_{\langle ij\rangle}=\pm1$ within the unit cell, where $\langle  ij\rangle$ are nearest neighbor bonds.  That $Q_{ij}$ is a pure phase results from the scale invariance introduced earlier. The restriction to Ising states then follows from the geometric frustration of the lattice and that $Q_{ij}$ is a pair amplitude which breaks the $U(1)$ gauge invariance down to $Z_2$\cite{Sachdev:1992qy}. We will not consider translationally non-invariant spin liquid states. This may be justified because the nature of the frustration of the hyper-kagome lattice, being made of corner-sharing triangles, is similar to the kagome lattice where the Sp($N$) ground state is translationally invariant\cite{Sachdev:1992qy}. 

%For example, in both cases coplaner spin configurations are favored and degeneracy is lifted at a finite size loop within a translationally invariant ansatz (as discussed below).

The search for spin liquid states is then simplified by performing a symmetry analysis on the above set of mean field states. By fixing the gauge, the symmetry of each state is then understood by studying its distribution of flux. The central result of this analysis\cite{inprep} is that two Z$_2$ spin liquid states \emph{and} their 8-fold degenerate topological sectors are identified. The 8-fold degeneracy arises from the existence of $\pi$ flux through loops that go all the way around the system, and is expected when periodic boundary conditions are imposed on a three-dimensional Z$_2$ spin liquid\cite{Wen:2004em}. We find that these two states are distinguished from each other by the existence or absence of $\pi$ flux through the 10 site loop (the smallest even site loop on this lattice).  We will therefore call them the zero-flux state and the $\pi$-flux state. 

\begin{figure}[t]
\subfigure[ordered zero-flux state]{\includegraphics[width=0.24\textwidth]{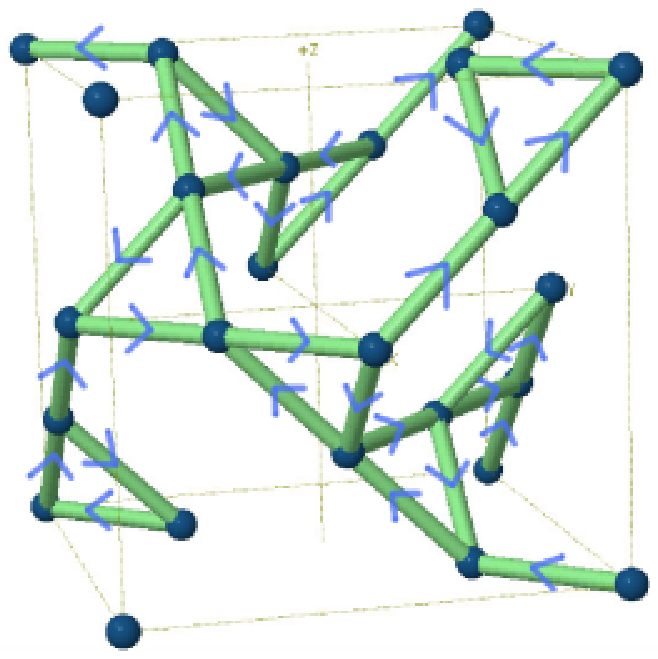}}
\subfigure[ordered $\pi$-flux state]{\includegraphics[width=0.23\textwidth]{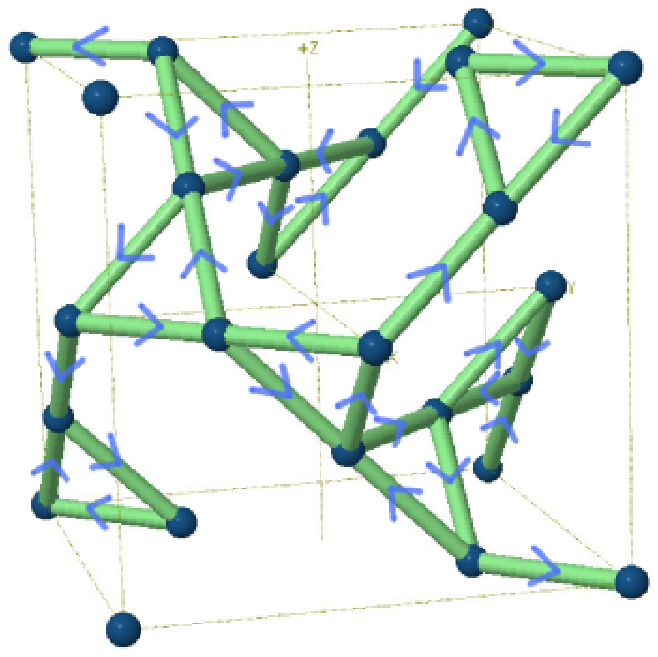}}
\caption{The zero- and $\pi$-flux states of the hyper-kagome Sp($N$) quantum antiferromagnet. If an arrow points from site $i$ to site $j$, then $Q_{ij}=+1$ and $Q_{ji}=-1$. }
\label{fig:hk-twostates}
\end{figure}

\emph{The zero-flux state is the mean field ground state.}  This follows from the flux expulsion principle. In addition, we can show this explicitly by making use of Eq. \ref{eq:deltaE} with $P_5=18264$ in the zero-flux state and $P_5=18224$ in the $\pi$-flux state. The energy difference at small $\kappa$ is then
\begin{equation}
  E_{\text{Sp($N$)}}[\pi\text{-flux}] - E_{\text{Sp($N$)}}[0\text{-flux}]\!=\!
      \frac{35}{2048}\kappa^5\! +\! O(\kappa^{6}) >0
\end{equation}
The zero-flux state has zero flux through all of its smallest loops and is unique (up to topological degeneracy) in the translationally invariant sector. Note, the splitting between these states is small in the limit of small $\kappa$, a consequence of the long (length 10) loop that needs to be traversed to induce a splitting. 

%This energy scale controls the $\pi$ flux (vison) line tension---hence low energy vison loops may be  expected in this geometry, which would correspond to low energy singlet excitations.

\emph{Topological order}. The topological order of the zero flux state, a central property of Z$_2$ spin liquids, requires an eight fold degenerate ground state manifold with periodic boundary conditions. In our construction, these eight degenerate states arise naturally.  Consider, for example, the zero-flux state with an additional $\pi$ flux through all $z$-axis ``threads"  (as highlighted in Fig. \ref{fig:hk-unitcell}) which wind around the system with $L$ unit cells in the $z$-direction ($L$ must be odd here). Since this state is otherwise the same as the state with zero flux through all threads, using Eq. \eqref{eq:deltaE} the energy splitting is
\begin{equation}\label{eq:Evison}
  \Delta E = \frac{2J}{3\sqrt{\pi}L^{3/2}}e^{-2L\ln (2/\kappa)}\! +\! O(\kappa^{2L+1})
\end{equation}
where $(2L-3)!!/(2L)!!\approx 1/\left(2\sqrt{\pi}L^{3/2}\right)$ for large $L$ and, since there are only two loops of length $2L$ starting from the same site, $P_{2L}^{n_z=0} - P_{2L}^{n_z=1}=4/3$. Note the exponential dependence of $\Delta E$ on $L$ is an explicit demonstration of the topological order of this phase\cite{Senthil:2001uc}. Furthermore, since this is a three-dimensional $Z_2$ spin liquid, a finite temperature transition to a paramagnetic state will occur in the inverted Ising universality class. This transition, however, is not in the spin sector and may not coincide with a cross-over in the spin structure factor from the high temperature CCP phase.

\emph{Experimental signatures.} The zero flux state has several features which distinguishes it from the CCP phase. Consider the energy integrated spin structure factor for the zero flux state shown in Fig. \ref{fig:hkSk}a (this was calculated following Ref. \onlinecite{Sachdev:1992qy}) and the static spin structure of the CCP phase (taken from Ref. \onlinecite{Hopkinson:2006qm}) shown in Fig.\ref{fig:hkSk}b. Notice that the spin liquid phase does not possess the long range dipolar spin correlations and its associated nodal structure along the $[hhh]$ direction (and symmetry-related directions)---characteristic features of the CCP phase. The presence/absence of long range dipolar correlations clearly distinguishes these two phases.

In the $Z_2$ spin liquid phase, the elementary excitations are spin-$1/2$ carrying spinions. Since the spin structure factor $S(\vec k,\omega)$ measures spin-$1$ excitations, it vanishes unless $\omega > \varepsilon(\vec q)+\varepsilon(\vec q-\vec k)$ for any value of $\vec q$ where $\varepsilon(\vec q)$ is the dispersion of a single spinon excitation. This defines the spinon-antispinon threshold\cite{Chung:2001th}. It is plotted as the bottom of the continuum given by the shaded (blue) region in Fig. \ref{fig:hkSk}c. It should be entirely absent in the CCP phase. 

\emph{Semi-classical regime.} Beyond the small  $\kappa$ regime, the zero-flux 
state becomes unstable to magnetic ordering at $\kappa=\kappa_c=0.4$ (similarly the $\pi$-flux state becomes unstable at $\kappa_c=0.8$).  This is remarkably large, given that magnetic order is expected to be more stable in three dimensions but that $\kappa_c=0.34$ for a triangular lattice and $\kappa_c=0.53$ for the kagome lattice\cite{Sachdev:1992qy}. The spin ordering pattern obtained upon spinon condensation in the zero-flux state is shown in Fig. \ref{fig:hk-unitcell}a. In contrast to $\sqrt{3}\times\sqrt{3}$ state on the kagome lattice, this state has no local weather-vane modes\cite{Chalker:1992fk}. Fluctuations within the classical spin manifold can only occur \emph{along an infinitely long thread} with pattern $BCBC$.  Thus, finite temperature fluctuations may not select this state consistent with the spin nematic ordering and the absence of magnetic ordering found in numerics\cite{Hopkinson:2006qm}.

\emph{Conclusion.} In this letter, we have presented tests for the existence of a quantum spin liquid phase in NIO. Using the large-$N$ Sp($N$) method, we proposed two candidate ground states: a magnetically ordered state in the semi-classical regime and a $Z_2$ quantum spin liquid state in a quantum regime. The magnetically ordered state has a $\vec q=0$ coplaner spin ordering pattern which we argue is unlikely to be entropically selected. Assuming NIO remains paramagnetic at low temperatures, it is still important to find features distinguishing a quantum spin liquid from a classical cooperative paramagnet (CCP). Here we have discussed two such features for the $Z_2$ spin liquid state. It should not have long ranged dipolar correlations and the associated nodal
structures along the $[hhh]$ direction (and symmetry related directions) that are the characteristics of the CCP phase. It should have a spinon-antispinon continuum with a threshold obeying a well defined dispersion.  Future neutron scattering experiments on NIO at current temperatures and at lower temperatures, would be decisive tests for our predictions.

\begin{acknowledgments}
We especially thank A. Paramekanti and H. Takagi for
many enlightening discussions. We also thank  J.-S. Bernier, J. Hopkinson, Y.-J. Kim, S.-H. Lee, R. Moessner, O. Tchernyshyov, M. Walker, F. Wang for useful
discussions. This work was supported by NSERC, CIAR, CRC (MJL, HYK,
YBK); KRF-2005-070-C00044, Visiting Miller Professorship at Berkeley
(YBK); the Hellman Family Faculty Award and LBNL DOE-504108 (AV).
\end{acknowledgments}

%\bibliography{hk-short}
%\bibliographystyle{apsrev}

\end{document}